\def\lsim{\mathrel{\rlap{\lower 4pt \hbox{\hskip 1pt $\sim$}}\raise 1pt \hbox
        {$<$}}}
\def\gsim{\mathrel{\rlap{\lower 4pt \hbox{\hskip 1pt $\sim$}}\raise 1pt \hbox
        {$>$}}}

\documentclass[12pt,preprint]{aastex}

\shorttitle{Nucleosynthesis in High-Entropy Hot-Bubbles of SNe}
\shortauthors{Izutani $\&$ Umeda}

\begin{document}


\title{Nucleosynthesis in High-Entropy Hot-Bubbles of SNe and
Abundance Patterns of Extremely Metal-Poor Stars}

\author{Natsuko Izutani$^{1}$ and Hideyuki Umeda$^{1}$}
\affil{$^{1}$Department of Astronomy, School of Science, University of Tokyo,
Hongo, Tokyo 113-0033, Japan}
\email{izutani@astron.s.u-tokyo.ac.jp; umeda@astron.s.u-tokyo.ac.jp}
\affil{Accepted to ApJL: July 21, 2010}

\begin{abstract}
There have been suggestions that 
the abundance of Extremely Metal-Poor (EMP) stars can be 
reproduced by
Hypernovae (HNe), not by normal supernovae (SNe).
However, recently it was also suggested that 
if the innermost neutron-rich or
proton-rich matter is ejected,
the abundance patterns of ejected matter are changed,
and normal SNe may also reproduce the observations of EMP stars.
In this letter, we calculate explosive nucleosynthesis with
various $Y_{\rm e}$ and entropy, and 
investigate whether normal SNe with this innermost matter,
which we call ``hot-bubble'' component,
can reproduce the abundance of EMP stars.
We find that neutron-rich ($Y_{\rm e}$ = 0.45-0.49) and 
proton-rich ($Y_{\rm e}$ = 0.51-0.55) matter
can increase Zn/Fe and Co/Fe ratios as observed,
but tend to overproduce other Fe-peak elements.

In addition to it, we find that if slightly proton-rich matter
with 0.50 $\le$ $Y_{\rm e}$ $<$ 0.501 with $s$/$k_{\rm b}$ $\sim$ 15-40
is ejected as much as $\sim$ 0.06 $M_\odot$, 
even normal SNe can reproduce the abundance of EMP stars,
though it requires fine-tuning of $Y_{\rm e}$.  
On the other hand, HNe can more easily reproduce the observations of
EMP stars without fine-tuning.
Our results imply that HNe are the most possible origin of 
the abundance pattern of EMP stars.

\end{abstract}
\section{Introduction}
The observational trends of extremely metal-poor (EMP) stars
reflect SN nucleosynthesis of Population (Pop) III,
or almost metal-free stars.
Their observed abundances show quite interesting patterns.
There are significant differences between the abundance patterns
in the iron-peak elements below and above
[Fe/H] $\sim$ -2.5.
For [Fe/H] $\lsim$ -2.5, the mean value of [Cr/Fe] and [Mn/Fe]
decrease toward lower metallicity, while [Co/Fe] and [Zn/Fe] increase
(McWilliam et al. 1995; Cayrel et al. 2004).

Umeda $\&$ Nomoto (2002, 2005) and Tominaga, Umeda $\&$ Nomoto (2007)
show that these trends can be related to
the variations of explosion energy of core-collapse SNe,
i.e, high [Zn,Co/Fe] and low [Cr,Mn/Fe] can be explained by 
a high energy SN ('hypernova'), while low [Zn,Co/Fe] and high [Cr,Mn/Fe]
by a normal SN.
On the other hand, Heger $\&$ Woosley (2008)
studied the evolution and parameterized explosions of Pop III
stars with masses ranging from 10 to 100 $M_\odot$. 
They have concluded that the EMP stars do not show the need for 
a HN component,
and that explosion energies less than 1.2 B seem to be preferred, though
their models tend to underproduce Co and Zn.

These previous works do not include any contribution from 
hot bubbles in the innermost region of SNe.
However, recent multi-dimensional simulations have shown that
both the neutron-rich and proton-rich matter in hot-bubbles
are ejected from the hot-bubble regions (e.g., Janka et al. 2003).
This innermost matter with various $Y_{\rm e}$ and entropy
is considered to be the origin of the heavier elements than 
Zn, but also can be the important site for the lighter elements
(e.g., Hoffman et al. 1996).
Heger $\&$ Woosley (2008) suggest that
Co and Zn in their models can be enhanced by this innermost matter.

Our previous work, Izutani, Umeda $\&$ Tominaga (2009) have studied
nucleosynthesis of
lighter neutron-capture elements 
(weak r-process elements), Sr, Y, and Zr
by considering small amount of mass ejection
from below the conventional mass-cut,
or from the hot-bubble regions.
In that paper, we assumed that the ``hot-bubble'' matter has
the same entropy with the supernova shock,
though in reality it may have higher entropy 
(e.g., in Janka et al. 2003 the hot-bubble matter has the entropy in the range,
$s$/$k_{\rm b}$ $\sim$ 30-50).
We have found that HN models with neutron-rich matter can reproduce 
these elements, but normal SN models cannot 
when we assume the same entropy with the supernova shock
for the matter below the mass-cut.
However, from the observational trends of [Zn/Fe] and [Sr/Fe]
(see Figure 15 and Discussion in Izutani et al. 2009),
it is highly possible that the main origin of 
the weak r-process is normal SNe.
In addition to it, some weak r-process stars have also
Mo, Ru, and Rh, though HN model cannot
reproduce those elements.
Therefore, we need to consider higher-entropy
calculations to clarify
the origin of these elements.

In this letter, we perform similar calculations 
as Izutani et al. (2009)
with wider range of $Y_{\rm e}$ and entropy.
We discuss whether normal SNe with the high-entropy
hot-bubble component can reproduce
the observations of EMP stars
especially paying attention to Fe-peak elements.
We also discuss some implications for the origin of 
the weak r-process elements, Sr-Rh.

\section{MODEL $\&$ METHOD}
The calculation method and other assumptions
are the same as described in Umeda $\&$ Nomoto (2002, 2005)
and Izutani et al. (2009)
except for the size of the nuclear reaction networks.
In this paper, we adopt the Pop III progenitors 
as in Umeda $\&$ Nomoto (2002, 2005) and apply the model with
$M$ = 15 $M_\odot$ and $E_{51}$ = 1 
(normal SN model)
and $M$ = 25 $M_\odot$ and $E_{51}$ = 20 
(HN model).
Detailed nucleosynthesis is calculated as a
postprocessing after the hydrodynamical calculation
with a simple $\alpha$-network.
The isotopes included in the post-process
calculations for $Y_{\rm e}$ $<$ 0.51 are 809 species up to $^{121}$Pd, 
and the ones for $Y_{\rm e}$
$\ge$ 0.51 are 652 species up to $^{112}$Pd.
(for detail, see Izutani, Umeda $\&$ Yoshida (2010) in preparation).
As for the HN model, 
we consider a model for which the density in the complete Si-burning region is
artificially reduced to 1/3 (or
entropy is enhanced by a factor of 3) of the original
as well as the original model.
The HN model with the entropy enhanced by a factor of 3 
fits the observation from Ca to Zn except Cr as shown in Figure 1.
We call this model ``HN3 model'' ($s$/$k_{\rm b}$ $\sim$ 50),
and call the HN model with the original entropy ``HN model'' ($s$/$k_{\rm b}$ $\sim$ 15).
As for the normal SN model, the density for the postprocessing
is multiplied by factors of 1 ($s$/$k_{\rm b}$ $\sim$ 5), 
1/3 ($s$/$k_{\rm b}$ $\sim$ 15), 1/10 ($s$/$k_{\rm b}$ $\sim$ 40), 
and 1/35 ($s$/$k_{\rm b}$ $\sim$ 150).
This modification of density mimics hot-bubbles in multi-dimensional simulations.

We obtain the final yields by adding matter above $M_{\rm cut}$
and below $M_{\rm cut}$.     
We set $M_{\rm cut}$ = 1.50 $M_\odot$ for
the normal SN model and 
1.88 $M_\odot$ for the HN3 model.
As a result, the normal SN model ejects 0.07 $M_\odot$ of
$^{56}$Ni, which is similar to SN1987A,
and the HN3 model ejects 0.51 $M_\odot$ of $^{56}$Ni.
\footnote{In the models assuming mixing-fallback effects,
the actual amount of the ejected $^{56}$Ni mass is smaller than this
value. See Umeda $\&$ Nomoto (2005).}

\section{Abundance Patterns of Whole Ejecta with Mass Ejection below
$M_{\rm cut}$}

In Figure 1, the abundance patterns from Si to Ru
are compared with those of EMP stars.
As for Co and Zn, the HN3 model reproduces the observations well,
while the normal SN model does not
as described in the figure caption.
As for Sr, Y, and Zr, both the HN3 model with
neutron-rich hot-bubble matter ($s$/$k_{\rm b}$ $\sim$ 15)
and the normal SN model with neutron-rich hot-bubble matter 
reproduce the observations well.
However, as for the heavier elements, Mo, Ru, and Rh,
only the normal SN model with $s$/$k_{\rm b}$ $\sim$ 150 neutron-rich
hot-bubble matter reach the observations.

\section{Parameter Dependence of Nucleosynthesis}

In this section, we show parameter dependences of the products
in hot-bubbles.
In Figure 2, we show [Co,Zn/Fe] vs $Y_{\rm e}$.
We show only the cases with 0.50 $\le$ $Y_{\rm e}$ $\le$ 0.501
in Figure 2 because
matter with $Y_{\rm e}$ in this range
shows better fitting to the abundance of EMP stars
than neutron-rich (0.45 $\le$ $Y_{\rm e}$ $\le$ 0.49)
and proton-rich (0.51 $\le$ $Y_{\rm e}$ $\le$ 0.55) matter
as mentioned in the next paragraph.
[Co/Fe] once increases as $Y_{\rm e}$ increases,
but decreases for larger $Y_{\rm e}$ in all the SN models.
$Y_{\rm e}$ at the [Co/Fe] peak becomes larger
with higher entropy.
[Co/Fe] has the peak at $Y_{\rm e}$ = 0.5003,
0.5004, and 0.5005 for $s$/$k_{\rm b}$ $\sim$ 5,
15, and 40, respectively.
[Co/Fe] becomes higher with higher entropy.
The highest [Co/Fe] is $\sim$ 0.5, 0.7, and 0.9
for $s$/$k_{\rm b}$ $\sim$ 5,
15, and 40, respectively.
[Zn/Fe] is nearly 0 in regardless of $Y_{\rm e}$
when $s$/$k_{\rm b}$ $\sim$ 5,
but decreases as $Y_{\rm e}$ increases 
in the higher-entropy ($s$/$k_{\rm b}$ $\sim$ 15 and 40) cases.
[Zn/Fe] becomes higher with higher entropy.
For example, [Zn/Fe] is $\sim$ 0.1, 0.4, and 0.9 with $Y_{\rm e}$ = 0.50
for $s$/$k_{\rm b}$ $\sim$ 5, 15, and 40, respectively.
For comparison, we also show [Zn,Co/Fe] in the complete Si-burning region
of the HN3 model by the horizontal lines.

Neutron-rich matter (0.45 $\le$ $Y_{\rm e}$ $\le$ 0.49)
also produces Co and Zn to some extent,
but at the same time, tends to overproduce Ni 
(Figure 3).
EMP stars show [Co/Ni] $\gsim$ 0, while
neutron-rich matter shows [Co/Ni] $<$ 0.
Proton-rich low-entropy matter also produces Co and Zn,
but the abundance of Co is smaller than
the case with
0.50 $\le$ $Y_{\rm e}$ $\le$ 0.501.
Proton-rich high-entropy matter produces less Co and Zn 
than proton-rich low-entropy matter.
In addition to it, both neutron- and proton-rich matter
tend to overproduce Cu.

\section{Trends of Fe-peak Elements of Whole Ejecta}
Here, we describe in detail the abundance patterns of
Fe-peak elements in the whole ejecta with mass ejection 
below $M_{\rm cut}$ or with hot-bubble component.
For the matter below $M_{\rm cut}$, the yields are averaged for $Y_{\rm e}$
values ranging from 0.45 to 0.50 for neutron-rich matter,
and from 0.51 to 0.55 for proton-rich matter.
The entropy below $M_{\rm cut}$ is fixed.
The other parameter of our model is the mass of ejected matter from
the regions below $M_{\rm cut}$, $\Delta M$.
$\Delta M$  is added to the matter above 
$M_{\rm cut}$, and they are assumed to be ejected to the outer space all together.
 
Figure 3 shows comparison between observed [Co/Fe], [Ni/Fe] vs. [Zn/Fe]
and those in our models.
These ratios are the ones of the total ejecta.
Therefore, the trends of their values are similar with the ones seen in Figure
2, but on the other hand, 
their exact values are different between Figure 2 and Figure 3
because the total ejecta includes the incomplete Si-burning region
where Fe is produced, but Co and Zn are not.
For example, [Zn/Fe] is $\sim$ 0.6, and [Co/Fe] is $\sim$ 0.2
in the total ejecta of HN3 (open cyan triangle in Figure 3),
while [Zn/Fe] is $\sim$ 0.9, and [Co/Fe] is $\sim$ 0.4
in the complete Si-burning region (blue and magenta horizontal lines in Figure 2).

The gradient of [Co/Fe] vs. [Zn/Fe] in EMP stars
is larger than that of the normal SN models with hot-bubble component.
The normal SN models with neutron-rich (0.45 $\le$ $Y_{\rm e}$ $\le$ 0.50)
matter (indicated by blue circles)
show the best Co/Zn among all the normal SN models.
However, Figure 3 right panel shows that the normal SN models 
with neutron-rich matter
tend to overproduce Ni (see also the caption of Figure 3)
as also mentioned in Section 4,
though [Ni/Fe] in EMP stars with [Zn/Fe] = 0.2-0.3 is well reproduced
if neutron-rich matter is added a little ($\lsim$ 0.006 $M_\odot$).
On the other hand, the HN3 model is marginally consistent 
with the observed [Co/Fe], and reproduces the observed [Ni/Fe] well.
In Figure 3, we also plot the ratios for the SN models 
with $Y_{\rm e}$ = 0.50, 0.5004, and
0.501 with $s$/$k_{\rm b}$ $\sim$ 15 (red filled circles),
and $Y_{\rm e}$ = 0.50, 0.5005, and 0.502 with $s$/$k_{\rm b}$ $\sim$ 40
(magenta filled circles).
We show $Y_{\rm e}$ = 0.5004 for $s$/$k_{\rm b}$ $\sim$ 15 matter,
and $Y_{\rm e}$ = 0.5005 for $s$/$k_{\rm b}$ $\sim$ 40 matter 
because [Co/Fe] has its peak at this $Y_{\rm e}$ value
in each case.
When $s$/$k_{\rm b}$ $\sim$ 15, [Co/Fe] has its peak value ($\sim$ 0.4)
at $Y_{\rm e}$ = 0.5004, and [Co/Fe] decreases with higher $Y_{\rm e}$.
[Co/Fe] is $\sim$ -0.1 at $Y_{\rm e}$  $=$ 0.501.
The similar trend of [Co/Fe] and $Y_{\rm e}$ can be seen when 
$s$/$k_{\rm b}$ $\sim$ 40.
These trends are consistent with the ones shown in Figure 2.

As for the HN model and the HN3 model, we also show a version, 
where the complete Si-burning
region is set to be $Y_{\rm e}$ = 0.5001 
(see Umeda $\&$ Nomoto 2005).
In this version, [Co/Fe] becomes higher, and the HN models reproduce
the observations well.
For example, 
[Zn/Fe] is $\sim$ 0.6, and [Co/Fe] is $\sim$ 0.2 in the HN3 model with $Y_{\rm e}$
= 0.50 (cyan open triangle with letters ``HN3 0.50'' in Figure 3), while 
[Zn/Fe] is $\sim$ 0.6, and [Co/Fe] is $\sim$ 0.3 in the HN3 model with $Y_{\rm e}$
= 0.5001 (cyan open triangle with letters ``HN3 0.5001'' in Figure 3).

\section{Discussion}
In this letter we have confirmed that only hypernova type
mass ejection can reproduce abundance of Fe-peak elements in EMP stars,
and neither neutron- or proton-rich high-entropy matter
ejection can help.
This also means that 
if normal SNe eject hypernova-like matter below $M_{\rm cut}$, 
then even normal SNe can reproduce [Co,Zn/Fe] in EMP stars.
More quantitatively, we find that the observations require ejection of
as much as 0.06 $M_{\odot}$ matter with $Y_{\rm e}$ $=$ 0.500-0.501
with $s$/$k_{\rm b}$ $\sim$ 15-40.
This condition may be difficult to realize for a normal SN.
For example, according to Janka et al. (2003), in their models
the amount of $Y_{\rm e}$ $<$ 0.47, $\le$ 0.50, and $>$ 0.50
matter is $<$ 10$^{-4}$ $M_\odot$, 6$\times$10$^{-3}$ $M_\odot$, and 0.03
$M_\odot$,
respectively,
and only small fraction of matter is in the range of $Y_{\rm e}$ $=$
0.500-0.501.
Therefore, this solution seems too contrived,
though
SN explosion mechanism is still uncertain and thus we cannot conclusively deny
such a possibility.

Here we discuss the additional cases with the different timescales,
the time it takes the flow to cool from some temperature to lower temperature.
As a result, [Co/Fe] is changed by the timescale 
below $T$ = 4 $\times$ 10$^9$ K 
(Pruet et al. 2005).
The timescale below $T$ = 4 $\times$ 10$^9$ K 
multiplied by a factor of 1/2 and 2
gives [Co/Fe] $\sim$ 0.5 and 0.2, respectively,
while [Co/Fe] $\sim$ 0.4 in the case with the original timescale 
as mentioned in Section 5.
However, Ye range and ejected mass needed is not so changed; 
the upper limit of Ye becomes a little larger, Ye $\sim$ 0.502, 
and ejected mass, a little smaller $\sim$ 0.03 $M_\odot$.

There is another reason why SN models are not favored to
explain large [Co,Zn/Fe] stars.
As mentioned in Introduction, [Co/Fe] decreases with higher metallicity,
and many stars show [Co/Fe] $\sim$ 0 when [Fe/H] $\gsim$ -2.5.
At this age, only core-collapse SNe are considered to contribute
to interstellar matter from the observation of [$\alpha$-elements/Fe].
If the typical value of [Co/Fe] in normal SN is $\sim$ 0.3, 
[Co/Fe] does not decrease to $\sim$ 0 at [Fe/H] $\sim$ -2.5.
On the other hand, IMF average of the supernova ejecta
can be [Co,Zn/Fe] $\sim$ 0 even though HN models
have large [Co,Zn/Fe]
(Tominaga et al. 2007).

Therefore, we suggest that Co and Zn in EMP stars
originated from HNe, not from normal SNe.
As for weak r-process elements,
it is not clear that
Sr-Zr rich weak r-process stars 
are always Mo-Rh rich.
Observational data of both Sr-Zr and Mo-Rh
are obtained in only two weak r-process stars
HD122563 and HD88609 (Honda et al. 2007).
Sr-Zr are produced in neutron-rich hot-bubble of 
$s$/$k_{\rm b}$ $\sim$ 15, while
both Sr-Zr and Mo-Rh are produced in neutron-rich hot-bubble
of $s$/$k_{\rm b}$ $\sim$ 150,
that may
be the part of neutrino-driven wind, rather than hot-bubble component
(Roberts et al. 2010).
Of course, this work is only parametric search,
and some other possibilities may exist in reality.
Further investigation is needed to disclose the origin of 
the abundances in EMP stars.

\acknowledgments

This work was partly supported 
by Grants-in-Aid for JSPS Fellows.


\begin{figure}
\plotone{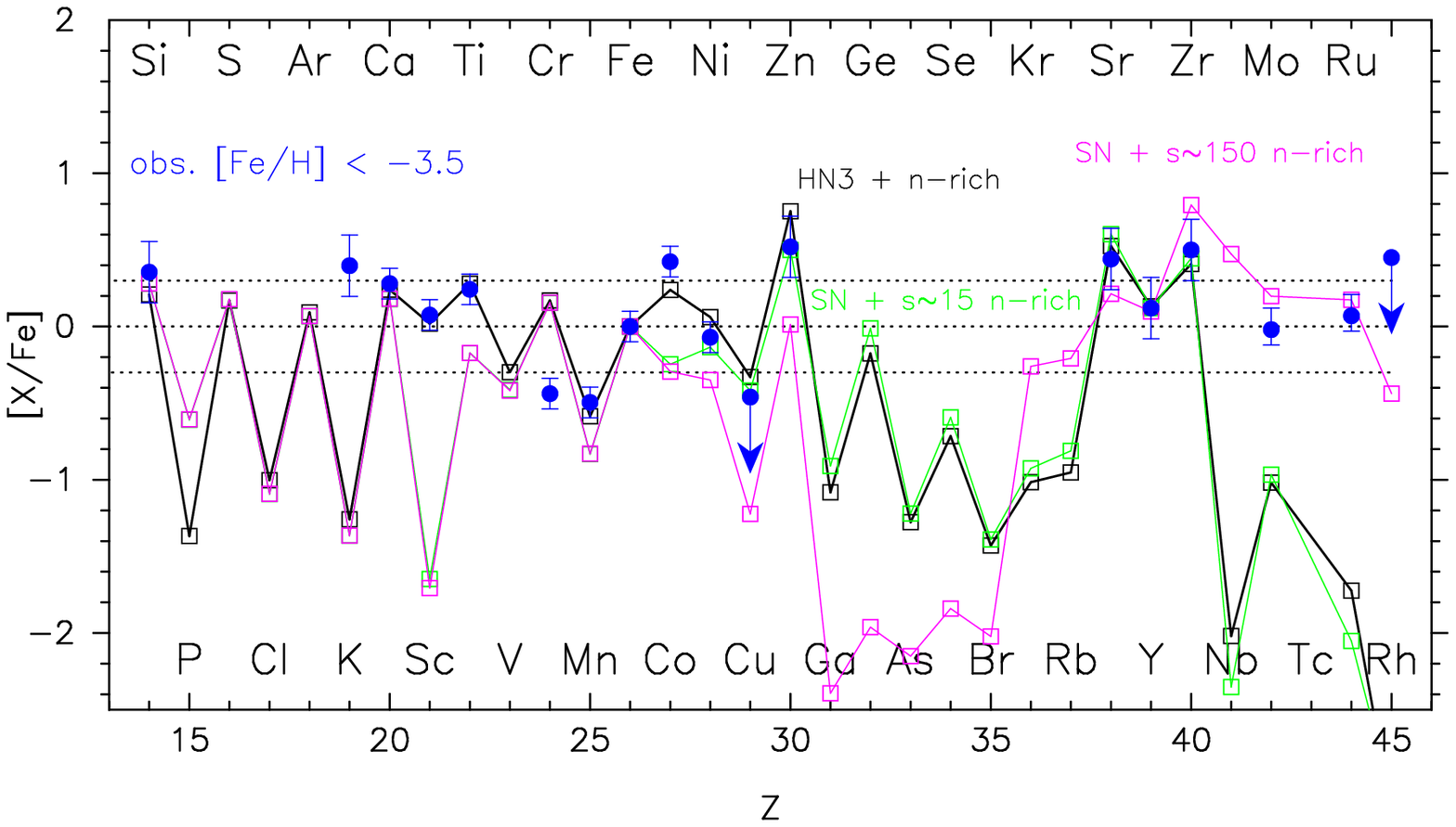}
\caption{Comparison between the yields of our models 
(HN3 model: black lines; SN model with $s$/$k_{\rm b}$ $\sim$ 15
neutron-rich matter: green lines; SN model
with $s$/$k_{\rm b}$ $\sim$ 150 neutron-rich matter: magenta lines)
and the abundance patterns of the EMP stars (blue circles with error bars).
The observations from Si to Zn are the average of four EMP stars
with -4.2 $<$ [Fe/H] $<$ -3.5 from Cayrel et al. (2004).
The observations from Sr to Zr, and from Mo to Rh are
the ones of CS 22897-008 (Fran\c cois et al. 2007) and HD 122563 
(Honda et al. 2007), respectively.
Among them, only the HN3 model reproduces both Co and Zn.
In addition to it, the HN3 model may reproduce Sr, Y, and Zr
if hot-bubble component is added as Izutani et al. (2009).
The normal SN model with $s$/$k_{\rm b}$ $\sim$ 15 neutron-rich matter
has high ratio of Zn, but Co is not good.
The normal SN model with $s$/$k_{\rm b}$ $\sim$ 150 neutron-rich matter
reproduces not only [Sr,Y,Zr/Fe] $\sim$ 0, but also 
[Mo,Ru,Rh/Fe] $\sim$ 0.
The models with proton-rich matter reproduce neither Co nor weak
r-process elements, which we do not show in this figure.
}
\end{figure}

\begin{figure}
\plotone{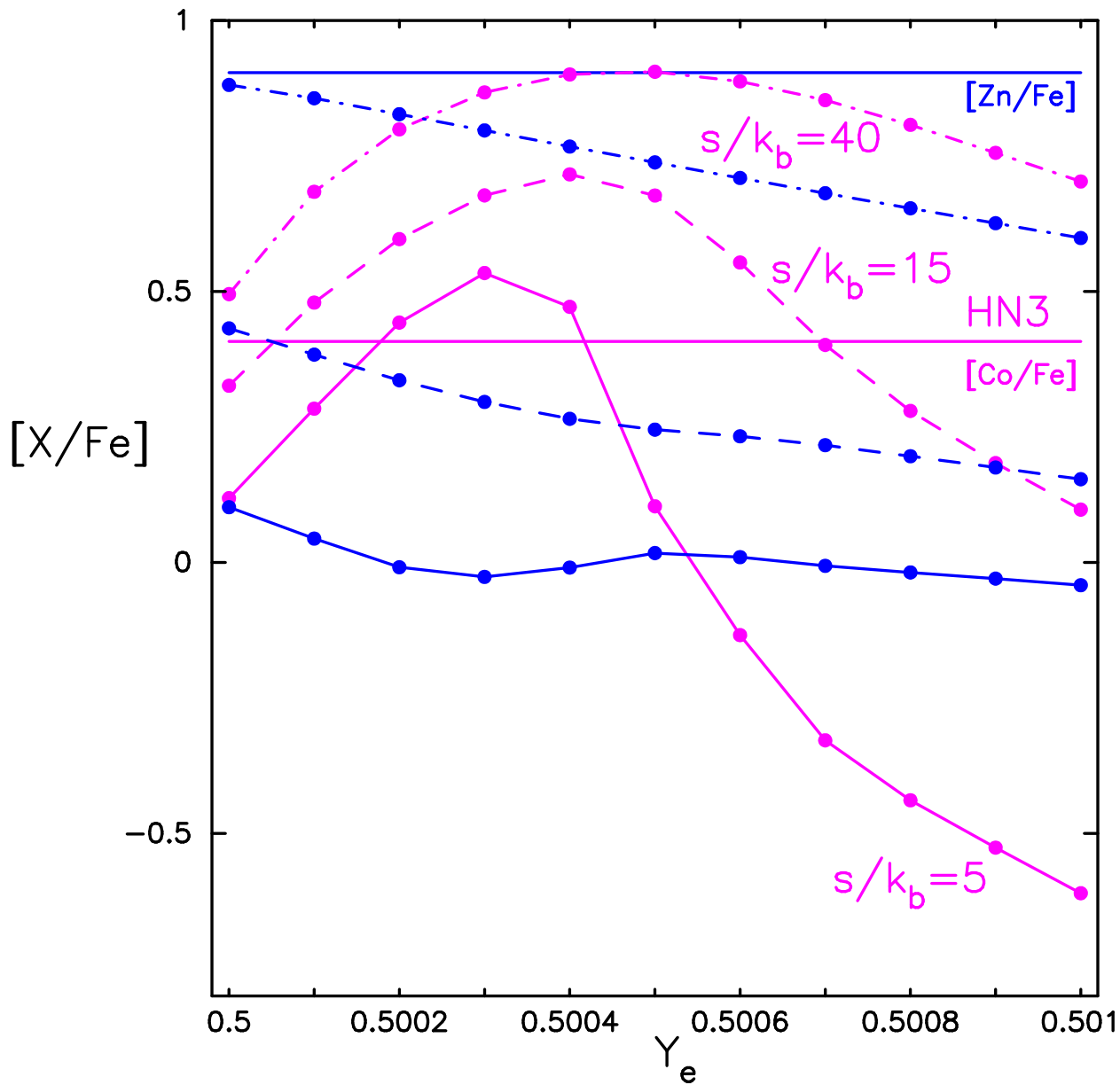}
\caption{[Co,Zn/Fe] in ``hot-bubbles'' of SNe.
[Co/Fe] and [Zn/Fe] are represented by magenta and blue figures,
respectively.
The ratios for $s$/$k_{\rm b}$ $\sim$ 5 (original), 15, and 40
flows in the normal SN model
are represented by solid, dashed, and dash-dotted lines, respectively.
The horizontal lines indicate the ratios in the HN3 model 
with the original $Y_{\rm e}$, 0.50.}
\end{figure}

\begin{figure}
\plotone{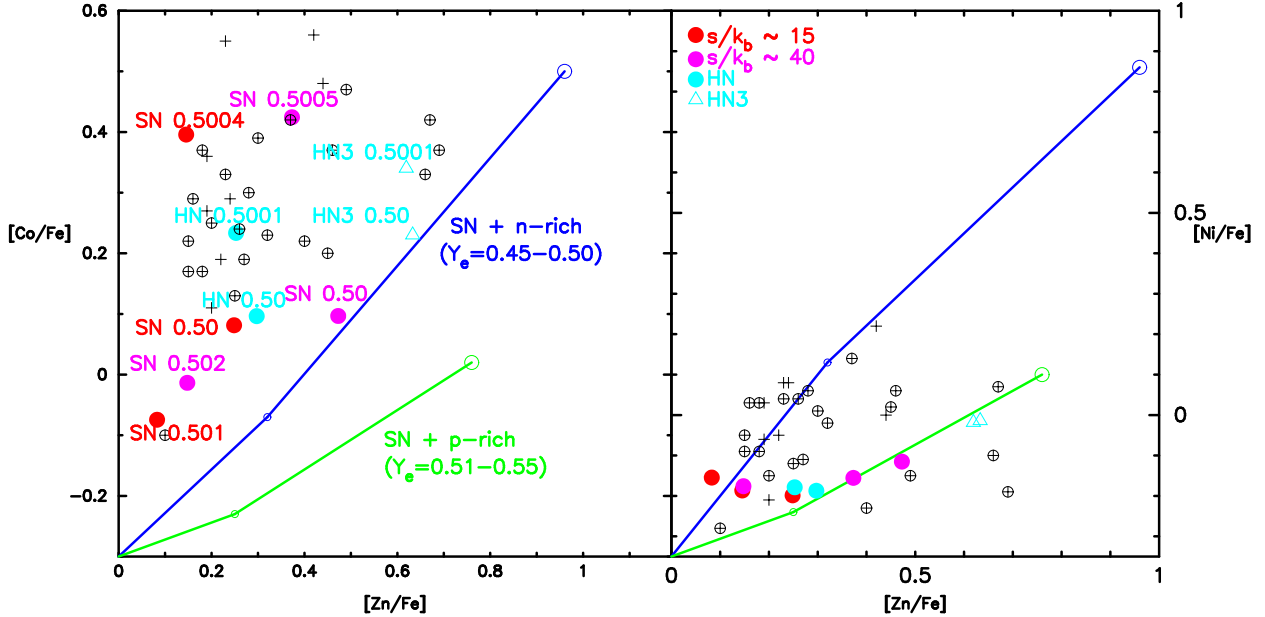}
\caption{Comparison between observed abundance ratios
and those in our models.
Left panel: [Co/Zn] vs. [Zn/Fe].
Right panel: [Ni/Fe] vs. [Zn/Fe].
The observations are represented by black crosses.
The observations with high [Sr/Ba] and [Y/Eu] 
(weak-r process star: see Izutani et al. 2009 for detail) are
enclosed by black circles.
The ratios for the normal SN model
with neutron-rich and proton-rich matter are represented by 
blue and green circles, respectively.
Small and big circles indicate the models with
$\Delta M$ = 0.006 $M_\odot$ and 0.06 $M_\odot$,
respectively.
For the SN models, we show only the original entropy models because
they have better [Co/Fe]
than higher entropy models.
All the SN models with a hot-bubble component
do not reach the observations of [Co/Fe].
Among them, the SN model with neutron-rich matter ($Y_{\rm e}$ = 0.45-0.50)
of $s$/$k_{\rm b}$ = 5 shows the best Co/Zn ([Co/Fe] $\sim$ 0.3
at [Zn/Fe] $\sim$ 0.7)
but tends to overproduce Ni ([Ni/Fe] $\sim$ 0.5
at [Zn/Fe] $\sim$ 0.7).
The ratios for the HN models and the HN3 models
are represented by cyan filled circles and open triangles, respectively.
We also show the ratios for the SN models
with $Y_{\rm e}$ = 0.50, 0.5005, and 0.502 with 
$s$/$k_{\rm b}$ = 40 (magenta filled circles),
and the ratios for the SN models
with $Y_{\rm e}$ = 0.50, 0.5004, and 0.501 with
$s$/$k_{\rm b}$ = 15 (red filled circles).
$\Delta M$ is 0.06 $M_\odot$ in these 6 models.
The ratio for the normal SN model without hot-bubbles 
is located at the coordinate origin.}
\end{figure}

\end{document}